\documentclass[useAMS,usenatbib]{mn2e}

\usepackage{epsfig}

\title[UV luminosity versus Ly$\alpha$ EW in high $z$ galaxies]{On the dependence between UV luminosity and Ly$\alpha$ equivalent width in high redshift galaxies}
\author[K.K. Nilsson, O. M{\"o}ller-Nilsson, P. M{\o}ller,  J.P.U. Fynbo \& A.E. Shapley]{K.K. Nilsson$^{1,2}$\thanks{E-mail:
knilsson@eso.org}, O. M{\"o}ller-Nilsson$^{2}$, P. M{\o}ller$^{3}$,  J.P.U. Fynbo$^{4}$ and A.E. Shapley$^{5}$\\
$^{1}$ST-ECF, Karl-Schwarzschild-Stra\ss e 2, 85748, Garching bei M\"unchen, Germany\\
$^{2}$Max-Planck-Institut f{\"u}r Astronomie, K{\"o}nigstuhl 17, 69117, Heidelberg, Germany\\
$^{3}$European Southern Observatory, Karl-Schwarzschild-Stra\ss e 2, 85748, Garching bei M\"unchen, Germany\\
$^{4}$Dark Cosmology Centre, Niels Bohr Institute, University of Copenhagen, Juliane Maries Vej 30, 2100 Copenhagen $\O$, Denmark\\
$^{5}$Department of Physics and Astronomy, 430 Portola Plaza, University of California, Los Angeles, CA 90095-1547, USA}
\begin{document}

\date{Accepted XXX xxxxxxx XX. Received XXXX xxxxxxx XX; in original form XXXX xxxxxxxxx XX}

\pagerange{\pageref{firstpage}--\pageref{lastpage}} \pubyear{2009}

\maketitle

\label{firstpage}

\begin{abstract}
We show that with the simple assumption of no correlation between the Ly$\alpha$ equivalent width and the UV luminosity of a galaxy, the observed distribution of high redshift galaxies in an equivalent width -- absolute UV magnitude plane can be reproduced. We further show that there is no dependence between Ly$\alpha$ equivalent width and Ly$\alpha$ luminosity in a sample of Ly$\alpha$ emitters. The test was expanded to Lyman-break galaxies and again no dependence was found. Simultaneously, we show that a recently proposed lack of large equivalent width, UV bright galaxies (Ando et al. 2006) can be explained by a simple observational effect, based on too small survey volumes.
\end{abstract}


\section{Introduction}
Two of the most common methods to find high redshift galaxies are based on detecting the Lyman-break or the Lyman-$\alpha$ emission-line of the galaxy. The Lyman-break technique, finding so-called Lyman-break galaxies (LBGs), has been extremely successful in gathering samples of galaxies in the redshift range $z \approx 3 - 6$ (e.g. Steidel et al. 1996, 1999, Pettini et al. 2001, Shapley et al. 2003, Bunker et al. 2004, Ouchi et al. 2004). In parallel, the search for high redshift galaxies by targeting the Ly$\alpha$ emission of star-forming galaxies with narrow-band imaging, finding so-called Ly$\alpha$ emitters (LAEs), has also been successful in the redshift range $z \approx 2.5 - 7$ (e.g. M{\o}ller \& Warren 1993, Cowie \& Hu 1998, Fynbo et al. 2002, Fynbo et al 2003, Matsuda et al. 2005, Malhotra et al. 2005, Shimasaku et al. 2006, Tapken et al. 2006, Kashikawa et al. 2006, Venemans et al. 2007, Finkelstein et al. 2007, Nilsson et al. 2007, 2009, Ouchi et al. 2008). In both types of surveys, two of the most robustly measured properties are those of flux in the rest-frame ultra-violet ($M_{UV}$) and the equivalent width of the Ly$\alpha$ line (EW). For Ly$\alpha$ emitters, these numbers are a direct product of the selection method. For Lyman-break galaxies, the Ly$\alpha$ EW is only constrained in samples with spectroscopic follow-up. 

Lately, it has been suggested that there is a lack of high redshift galaxies with large Ly$\alpha$ EWs and large UV fluxes. In Ando et al. (2006) a sample of Lyman-break galaxies at $z \sim 5$ and $z \sim 6$ were studied for their Ly$\alpha$ equivalent widths and rest-frame UV fluxes. Their results were also compared to Ly$\alpha$ emitters at similar redshifts and they reported a lack of UV-bright objects with high EWs. The proposed explanations by the authors were those of different dust extinctions, smaller/larger amount of neutral hydrogen gas, age differences or gas kinematic effects. However, Ando et al. (2006) also suggested that larger samples are necessary to confirm this apparent deficiency of UV-bright, high Ly$\alpha$ EW objects. Following this publication, several other authors have claimed to see the same effect. In the observational publications of Ando et al. (2007), Iwata et al. (2007), Ouchi et al. (2008), Vanzella et al. (2009) and Shioya et al. (2009) a lack of these objects is reported. Ando et al. (2007) and Vanzella et al. (2009) both suggest that this may be a result of larger amounts of dust and a higher metallicity in UV-bright galaxies. Ouchi et al. (2008) propose that it infers a young, low-mass population of galaxies at high redshift. The lack of high-EW, UV-bright galaxies has also been seen in theoretical work. Mao et al. (2007) present a simple physical model for high redshift galaxies, including a dust screen model. In parallel, Kobayashi et al. (2009) developed a semi-analytical model of high redshift galaxy evolution, including gas not homogeneously distributed, but rather in clumps. They both find their models to predict the lack of high-EW, UV-bright galaxies. On the other hand, there are also publications claiming not to see this effect. Verma et al. (2007) and Stanway et al. (2007) both study samples of $z = 5-6$ Lyman-break galaxies and see no UV dependence on the EW. Neither did Deharveng et al. (2008), observing nearly 100 Ly$\alpha$ emitters at $z \sim 0.3$. Deharveng et al. (2008) also argue that the claim to see such an effect may be due to small number statistics. This is the same argument as Dijkstra et al. (2007) propose. Their simple SFR-based galaxy evolution model does not reproduce any correlation between UV flux and Ly$\alpha$ EW. 

In two recent publications, Pflamm-Altenburg et al. (2009) and Meurer et al. (2009) discuss how a varying IMF can cause a decline in H$\alpha$ luminosity, and thus the EWs, at bright UV fluxes. 
Even though it is unclear what the relation between Ly$\alpha$ and H$\alpha$ is in the high redshift Universe, there should exist a correlation between the two fluxes, and hence 
this effect could potentially also cause the Ly$\alpha$ EW to decrease. Pflamm-Altenburg et al. (2009) predict that this effect will take place at star formation rates below $0.01$~M$_{\odot}$~yr$^{-1}$. As surveys for high redshift galaxies are unable to reach these low star formation rates, this particular effect is not expected to be observable. 

Given the large number of recent publications concerned with this
issue, the number of proposed interpretations, and given that
there still seems to be disagreement as to whether an effect is actually
observed, we find it appropriate to subject the issue to a rigorous
statistical test. In this paper we address \emph{if} there is a dependency between the Ly$\alpha$ EW and UV luminosity (or Ly$\alpha$ flux) in observed, high redshift galaxy samples. We focus on results from $z\sim3$ Ly$\alpha$ emitter, and Lyman-break galaxy, surveys as the amount of available, observational data is the largest at that redshift. Section~\ref{sec:model} describes the method used. The results are given in sec.~\ref{sec:results} and a discussion of the results is found in sec.~\ref{sec:discussion}.

\vskip 5mm Throughout this paper, we assume a cosmology with $H_0=72$
km s$^{-1}$ Mpc$^{-1}$ (Freedman et al. 2001), $\Omega _{\rm m}=0.3$ and
$\Omega _\Lambda=0.7$.

\section[]{Method}\label{sec:model}
\begin{figure*}
  \epsfig{file=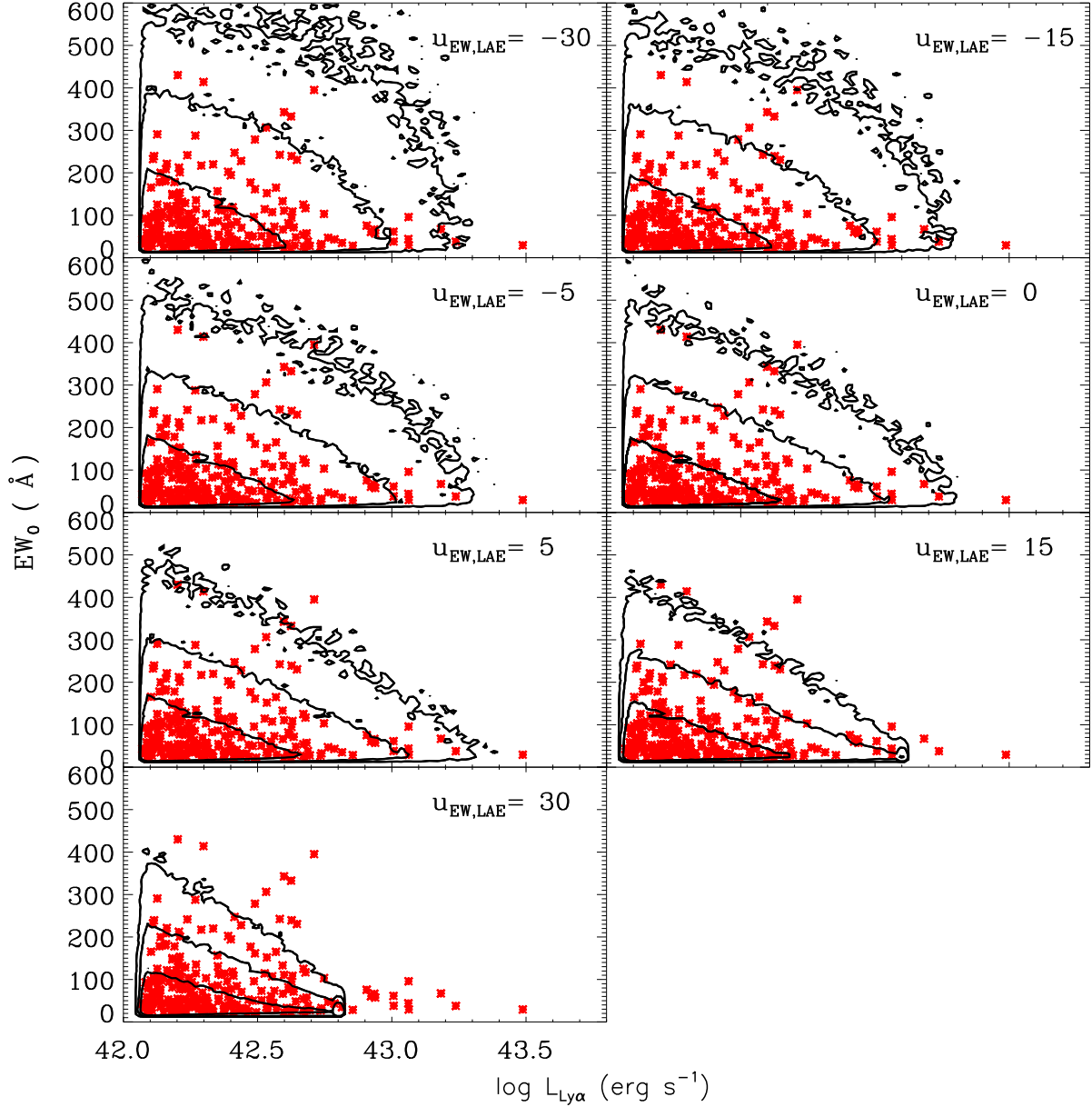,width=17.0cm}
 \caption{Plot of EW versus Ly$\alpha$ luminosity for Ly$\alpha$ emitters. Red stars mark the observed sample (see text for details). Each panel includes black contours with 68\%, 95\% and 99.7\% significance of the simulated sample for different values of $u_{EW,LAE}$ for LAEs. It is clearly seen that very large or very small $u_{EW, LAE}$ are ruled out by the observations.}
 \label{fig:ewuv}
\end{figure*}

The two observed parameters used here are the Ly$\alpha$ and UV fluxes of a galaxy. For simplicity, we choose to perform our tests on plots involving the EW of the galaxy (see also Fig.~\ref{fig:ewuv} or Fig.~\ref{fig:ando}) . This is a simple coordinate transformation, according to:
\begin{equation}\label{eq:ewdef}
EW_{Ly\alpha} \equiv \frac{F_{Ly\alpha}}{f_{UV}}
\end{equation}
where $F_{Ly\alpha}$ is the Ly$\alpha$ flux and $f_{UV}$ is the flux density in the UV at the Ly$\alpha$ wavelength. 
The 1-D EW distributions for high redshift galaxies, averaged over large samples, have been shown to closely resemble an exponentially declining function (Gronwall et al. 2007, Nilsson et al. 2009):  
\begin{equation}\label{eq:ewexp}
P(EW)dEW = cons. \times \exp(-EW/w_u) dEW
\end{equation}
Here the $w_u$ is a scaling constant that can be fitted for for each data-set. The suggestion that there may be a lack of high luminosity objects
with high EWs would then imply that $w_u$, at any given redshift,
must be a function
of the luminosity in the sense that the EW distribution becomes
narrower for high luminosity objects as the high EW tail of the
distribution is truncated. In principle this is easy to test simply by dividing a sample
of galaxies into luminosity bins and then ask if there is any
evidence for a significant change of $w_u$ between the high
and low luminosity sub-samples. In practice the expected number
of objects in the high luminosity, high EW bin is too small to
give a robust answer.

Instead, we introduce a potential dependence on the EW from the Ly$\alpha$/UV flux, allowing a Monte Carlo approach. To quantify the dependence, the parameter $u_{EW}$ is introduced into the scaling constant according to:
\begin{equation}\label{eq:ewlae}
w_{u,LAE} = w_{0,LAE} - u_{EW,LAE} \times \frac{\mathrm{L}_{Ly\alpha}}{10^{42.5}}
\end{equation} 
for Ly$\alpha$ emitters, where $w_{0,LAE}$ is the best fit scaling constant if $u_{EW,LAE} = 0$ and L$_{Ly\alpha}$ is the Ly$\alpha$ luminosity in erg~s$^{-1}$. For Lyman-break galaxies, instead we have: 
\begin{equation}\label{eq:ewlbg}
w_{u,LBG} = w_{0,LBG} - u_{EW,LBG} \times \frac{f_{UV}}{10^{-10.5}}
\end{equation} 
where 
\begin{equation}
f_{UV} = 10^{-0.4 \times (M_{UV}+48.6)}
\end{equation}
The constants in Eq.~\ref{eq:ewlae} and \ref{eq:ewlbg} are carefully chosen to allow a change in $w_u$ at the brightest end of the order $\pm w_0$ when $-30 \leq u_{EW} \leq 30$, and a change in the faintest end of the order a few {\AA}.
These equations resolve into no dependence between Ly$\alpha$/UV flux and EW if $u_{EW} = 0$ and a large dependence when $u_{EW} \to \sim \pm w_0$. A large positive $u_{EW}$ will result in fewer objects with large EWs at bright fluxes, and a large negative $u_{EW}$ in an excess of large EW objects at bright fluxes. Fig.~\ref{fig:ewuv} illustrates the effect of $u_{EW}$ on the EW-flux distribution.
It is important to note that it is necessary to fit the two types of galaxies with different equations, because they are flux limited in different parameters; Ly$\alpha$ emitters in Ly$\alpha$ flux and Lyman-break galaxies in the UV flux. Ideally, both galaxy samples would be fit with the same equation, with e.g. a dependency on only $f_{UV}$. This would, however, infer a flux dependent lower limit to the EW of the Ly$\alpha$ emitters, an effect which is clearly illustrated in the left panel of Fig.~\ref{fig:ando}. Because of this, we choose to fit Ly$\alpha$ emitters and Lyman-break galaxies with separate equations and dependencies. 

To test the value of $u_{EW}$, we make a simulated distribution of galaxies by drawing from 1-D distributions for the flux and the EW, with different values on $u_{EW}$, and test if the simulated galaxy distribution resembles the observed distribution. We do this in three steps. First we model the observed EW distribution, to find $w_0$. The procedure to find the best fit $u_{EW}$ then includes creating simulated EW$-$flux distributions of galaxies by drawing randomly from the Ly$\alpha$/UV luminosity functions and the exponentially declining EW distribution function according to Eq.~\ref{eq:ewexp} and~\ref{eq:ewlae} or~\ref{eq:ewlbg} for different values of $u_{EW}$. The 1-D flux distributions we draw from are the observed respective Ly$\alpha$ and UV luminosity functions, summarised in Table~\ref{tab:lfew}. As this paper is focused on results at redshift three, the luminosity function is drawn from Gronwall et al. (2007) in the case of Ly$\alpha$ emitters.  Several other luminosity functions have been published but they all agree very well (see also van Breukelen et al., 2005, Ouchi et al., 2008, Grove et al., 2009).  In the case of Lyman-break galaxies we use the luminosity function of Reddy et al. (2008).
\begin{table*}
 \centering
 \begin{minipage}{140mm}
  \caption{Luminosity functions and EW distributions used in the creation of a random simulated sample. }
  \begin{tabular}{@{}lccccccccc@{}}
  \hline
   Type & L$^*$ / M$^*$      &  $\phi^*$   &   $\alpha$   &    Type of EW dist.   &   $w_0$   &\\
              &   [erg~s$^{-1}$] / [---] & [Mpc$^{-3}$] &                &                                    & [{\AA}]   & \\
 \hline
 Ly$\alpha$ emitters & 42.66 & $1.28 \times 10^{-3}$ & $-1.36$ & Exp & $69$ & \\ 
 Lyman-break galaxies & $-21.12$ & $1.12 \times 10^{-3}$ & $-1.85$ & Exp & $29$ & \\ 
 \hline
\label{tab:lfew}
\end{tabular}
\begin{list}{}{}{}{}{}{}
\item[ ]  Luminosity function parameters are from Gronwall et al. (2007) for Ly$\alpha$ emitters and from Reddy et al. (2008) for Lyman-break galaxies. Widths of equivalent width distributions ($w_0$) are the one parameter fits to the total samples of Ly$\alpha$ emitters and Lyman-break galaxies as described in the text.
\end{list}
\end{minipage}
\end{table*}

Finally, we compare the simulated galaxies with the observed sample of 232 Ly$\alpha$ emitters and 128 Lyman-break galaxies (see sec.~\ref{sec:results} for a definition on the samples). For each simulated distribution, two statistical tests are performed to find the simulated distribution that best fits the observed data. Both methods are based on dividing the plane of data into four quadrants, with the intersection placed in an arbitrary point on the plane (see also Peacock 1983). In each quadrant the ratio between the number of galaxies in that quadrant to that of the number of galaxies in the total sample is computed. This is done for both the data sub-set sample and the simulated galaxy sample. This process is re-iterated for every point in the plane until the largest \emph{difference} between the observed and test sample ratios is found in one quadrant. In our case the sampling of the plane has effectively one million resolution elements ($1000\times1000$), ensuring that the maximum difference varies by less than $0.1$\% in the points nearest to the maximum.  Based on the largest difference found, the following two tests are performed:
\begin{enumerate}
\item Monte-Carlo test
\item 2-D Kolmogorov-Smirnov test
\end{enumerate}
In test \emph{(i)} we merely find the largest difference between the simulated sample and the data. We calculate the errors on the ratios of test and observed samples and can thus determine how dissimilar the two distributions are by calculating the difference in the ratios and the significance in the same. In test \emph{(ii)} we perform the 2-D Kolmogorov-Smirnov test as described in Peacock (1983). The significance of the two distributions being similar is determined by:
\begin{equation}\label{eq:KS}
P(>Z_\infty) = 2 \times \exp{\left(-2 (Z_\infty - 0.5)^2 \right)}
\end{equation}
\begin{equation}
Z_\infty = \frac{\sqrt{n} \times D_n}{1 - 0.53 n^{-0.9}}
\end{equation}
\begin{equation}
n = \frac{n_1n_2}{n_1+n_2}
\end{equation}
where $n_1$ and $n_2$ are the numbers of objects in the two respective samples and $D_n$ is the largest difference found in one quadrant. 
We end up with a set of simulated samples, one for each $u_{EW}$, each with a measure of how likely this sample resembles the real data. With these, we can find the distribution that best fits the data, and thus the best fit value of $u_{EW}$.

Doing this once on the total sample of data will only reveal the best fit $u_{EW}$, without any understanding of the biases in the observed sample, or on the error bars. To understand the full probability function of the $u_{EW}$ parameter from the data we employ a jack-knife technique to the observed data. For each data-set (Ly$\alpha$ emitters and Lyman-break galaxies) 3000 random sub-sets of data are created with the total number of galaxies in the total data-set minus ten. For each of these data sub-sets, a new $w_u$ has to first be fitted for, after which the previously explained analysis can be repeated. Thus, a best fit $u_{EW}$ is determined for each of these 3000 data sub-sets, allowing an analysis of the distribution of $u_{EW}$.

\section{Results}\label{sec:results}
\subsection{Ly$\alpha$ emitters}
\begin{figure*}
  \epsfig{file=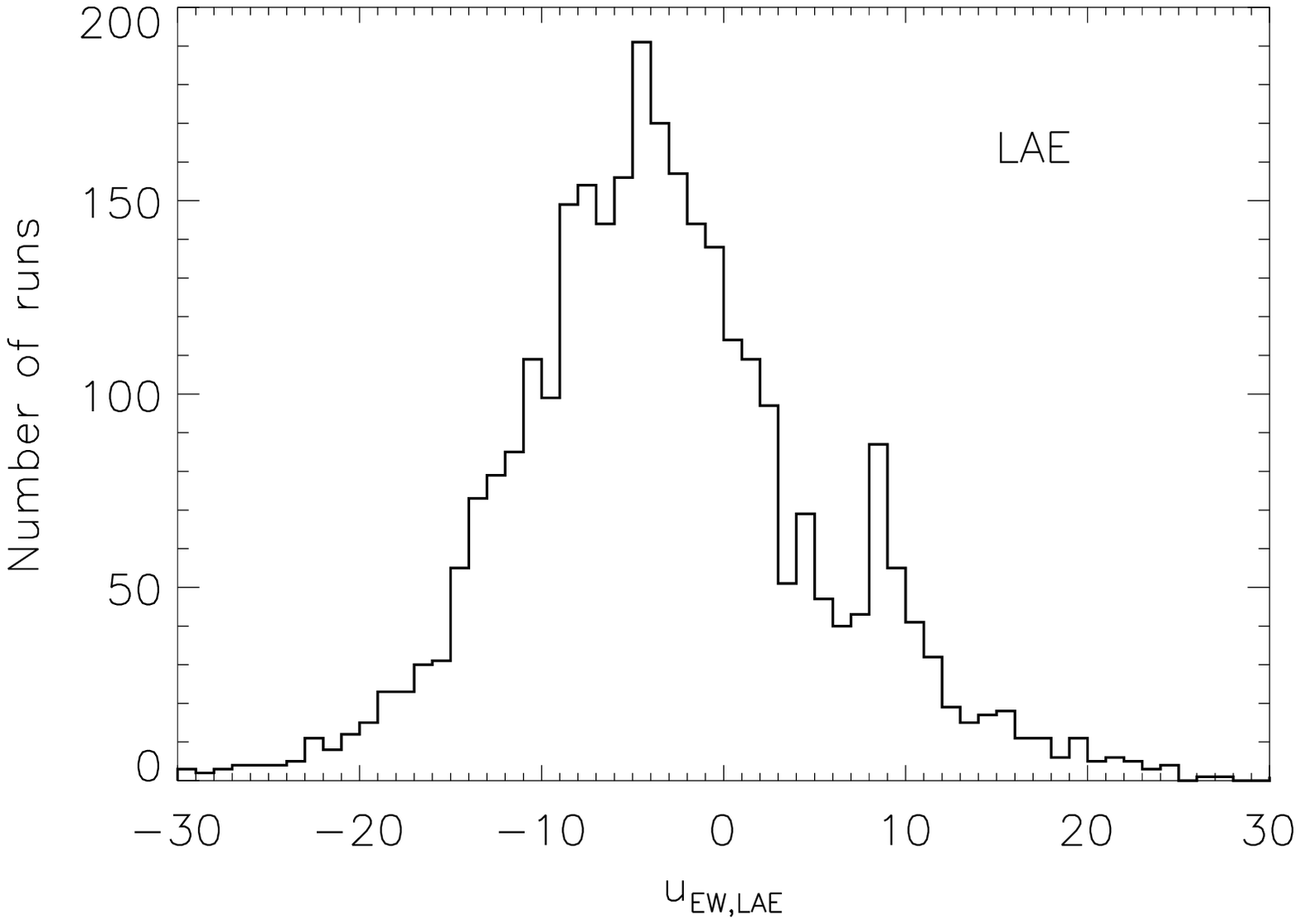,width=9.0cm}\epsfig{file=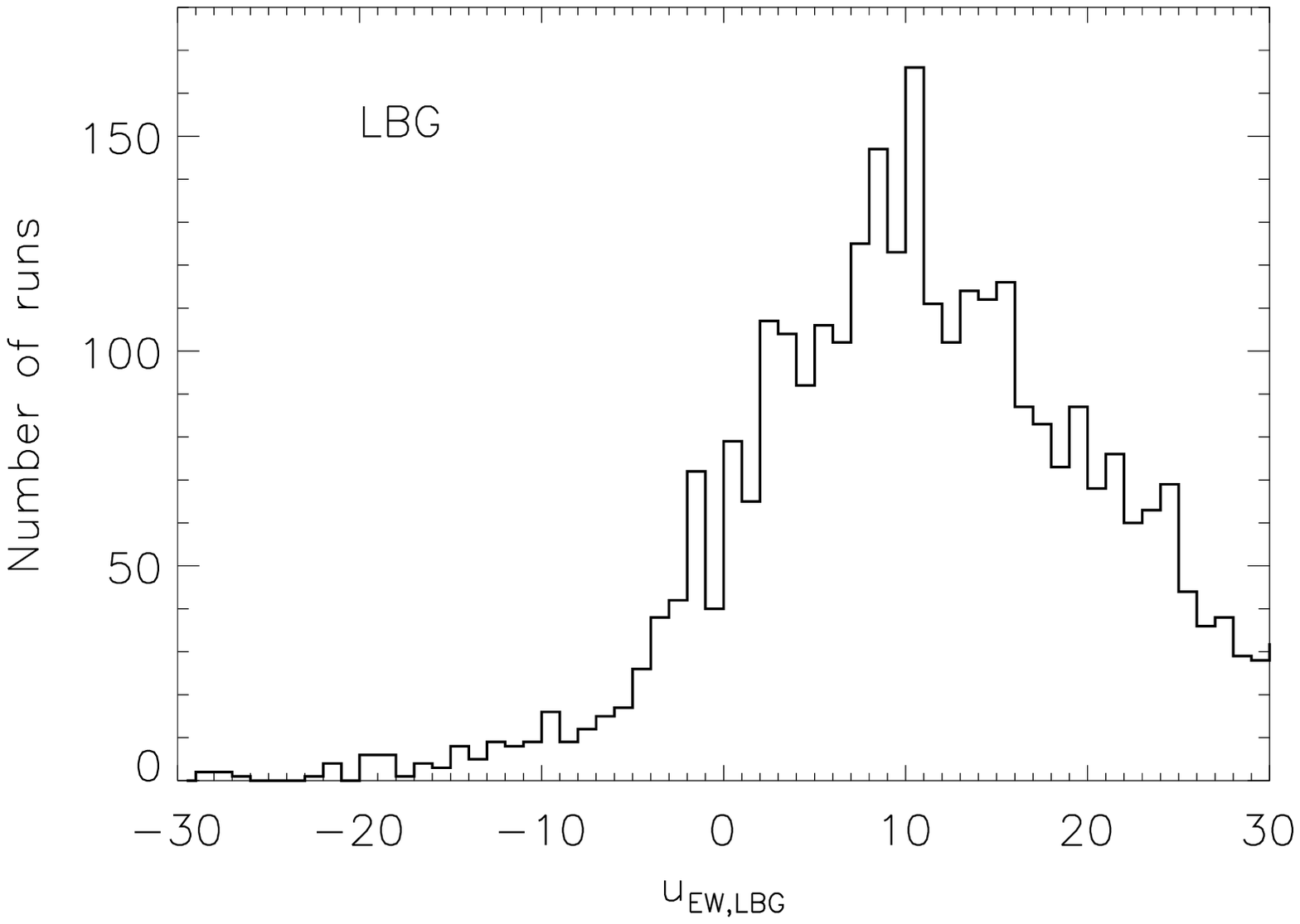,width=9.0cm}
 \caption{Histogram of best fit $u_{EW}$ for each of the 3000 random Ly$\alpha$ emitter and Lyman-break galaxy sub-sets. The best fit $u_{EW}$ are $u_{EW,LAE} = -4^{+9}_{-7}$ and $u_{EW,LBG} = 10^{+10}_{-9}$.}
 \label{fig:laeresults}
\end{figure*}

To test the dependence of EW and Ly$\alpha$ luminosity in Ly$\alpha$ emitters, the observed samples of $z \sim 3$ Ly$\alpha$ emitters of Gronwall et al. (2007), Nilsson et al. (2007), Venemans et al. (2007) and Grove et al. (2009) are used. A lower limit of restframe 20~{\AA} EW is set as this is typically the limiting EW for selection in narrow-band surveys. A lower limit to the Ly$\alpha$ luminosity is also set to be equal to that of Gronwall et al. (2007); $\log{\mathrm{L}_{Ly\alpha}} = 42.07$~erg~s$^{-1}$. After rejecting objects in these catalogues with EW$_0 < 20$~{\AA} and $\log{\mathrm{L}_{Ly\alpha}} < 42.07$~erg~s$^{-1}$, and known AGN, we are left with a total sample of 232 LAEs with recorded Ly$\alpha$ EW and luminosity. The $u_{EW,LAE}$ parameter is fit with values between $-30$ and $30$ in steps of one. In Fig.~\ref{fig:laeresults} the histogram of best fit $u_{EW,LAE}$ to each of the 3000 random data sub-sets is shown. 
The median of the distribution is $u_{EW,LAE} = -4$ and the mean is $u_{EW,LAE} = -3.66$. Based on the skewed appearance of the histogram, the error bars are calculated by separately integrating the histogram to 68.3~\% of each wing, split in the position of the median. The best fit $u_{EW,LAE}$ is then $u_{EW,LAE} = -4^{+9}_{-7}$.

\subsection{Lyman-break galaxies}
The largest sample of emission-line properties of Lyman-break galaxies has been published in Shapley et al. (2003). This is the sample that will be used here. An upper absolute UV magnitude is set to $-20.0$. EWs are constrained to be larger than $20$~{\AA} in order to ensure spectroscopic completeness. A total of 128 Lyman-break galaxies have been observed at redshifts $2.7 < z < 3.3$ with EW$_{0,Ly\alpha} > 20$ and M$_{UV} < -20.0$. This population of galaxies is fit by selecting an absolute UV magnitude from the luminosity function of Reddy et al. (2008, see also Table~\ref{tab:lfew}), and then an EW from an exponential function, see Eq.~\ref{eq:ewexp} and Eq.~\ref{eq:ewlbg}. The $u_{EW,LBG}$ parameter is fit with values between $-30$ and $30$ in steps of one. In Fig.~\ref{fig:laeresults} the histogram of best fit $u_{EW,LBG}$ to each of the 3000 random data sub-sets is shown. The median of the distribution is $u_{EW,LBG} = 10$ and the mean is $u_{EW,LBG} = 10.23$. The error bars are calculated in the same way as for Ly$\alpha$ emitters, resulting in $u_{EW,LBG} = 10^{+10}_{-9}$.

\subsection{On the proposed lack of high-EW, UV bright high redshift galaxies}
\begin{figure*}
  \epsfig{file=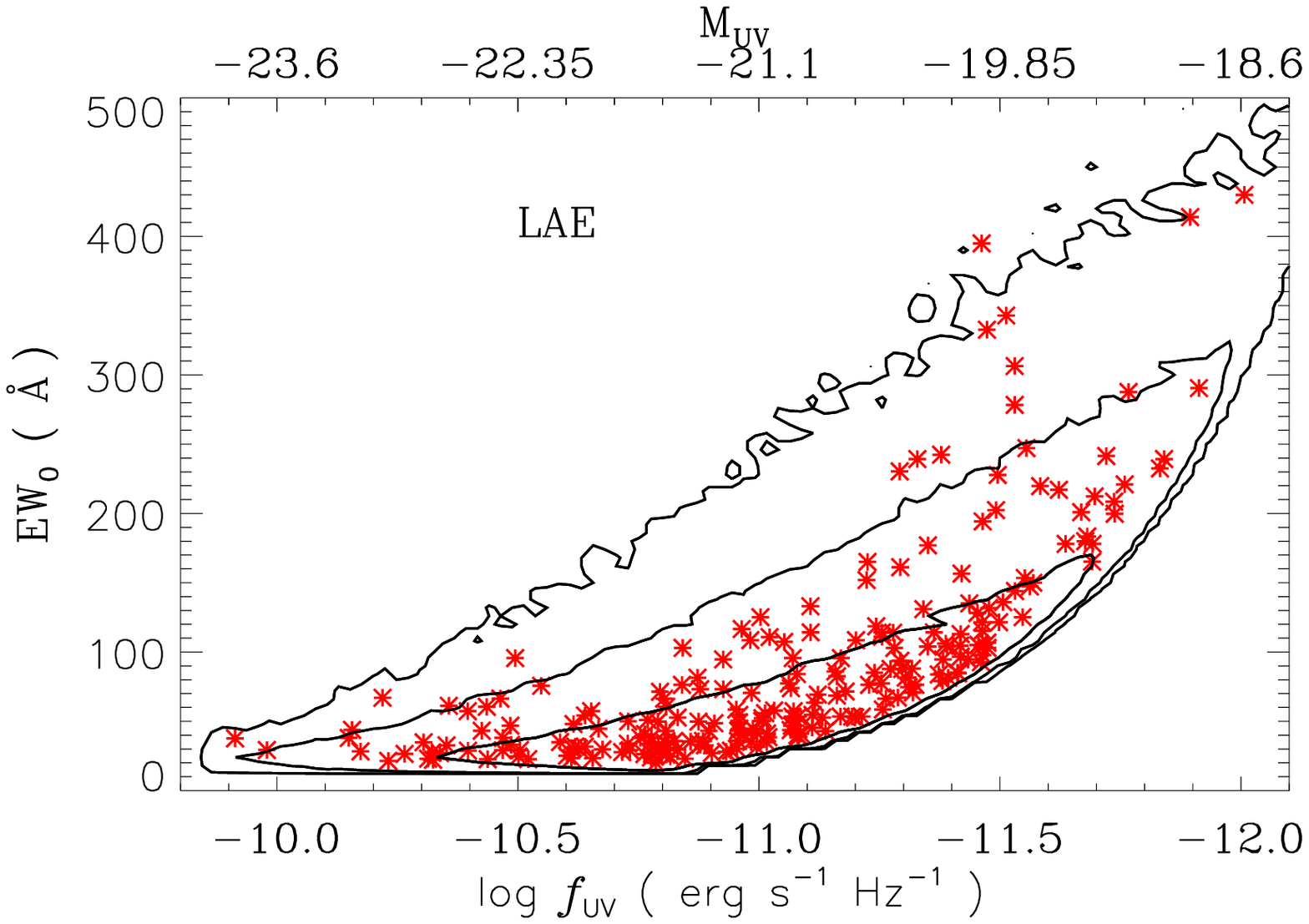,width=8.5cm} \epsfig{file=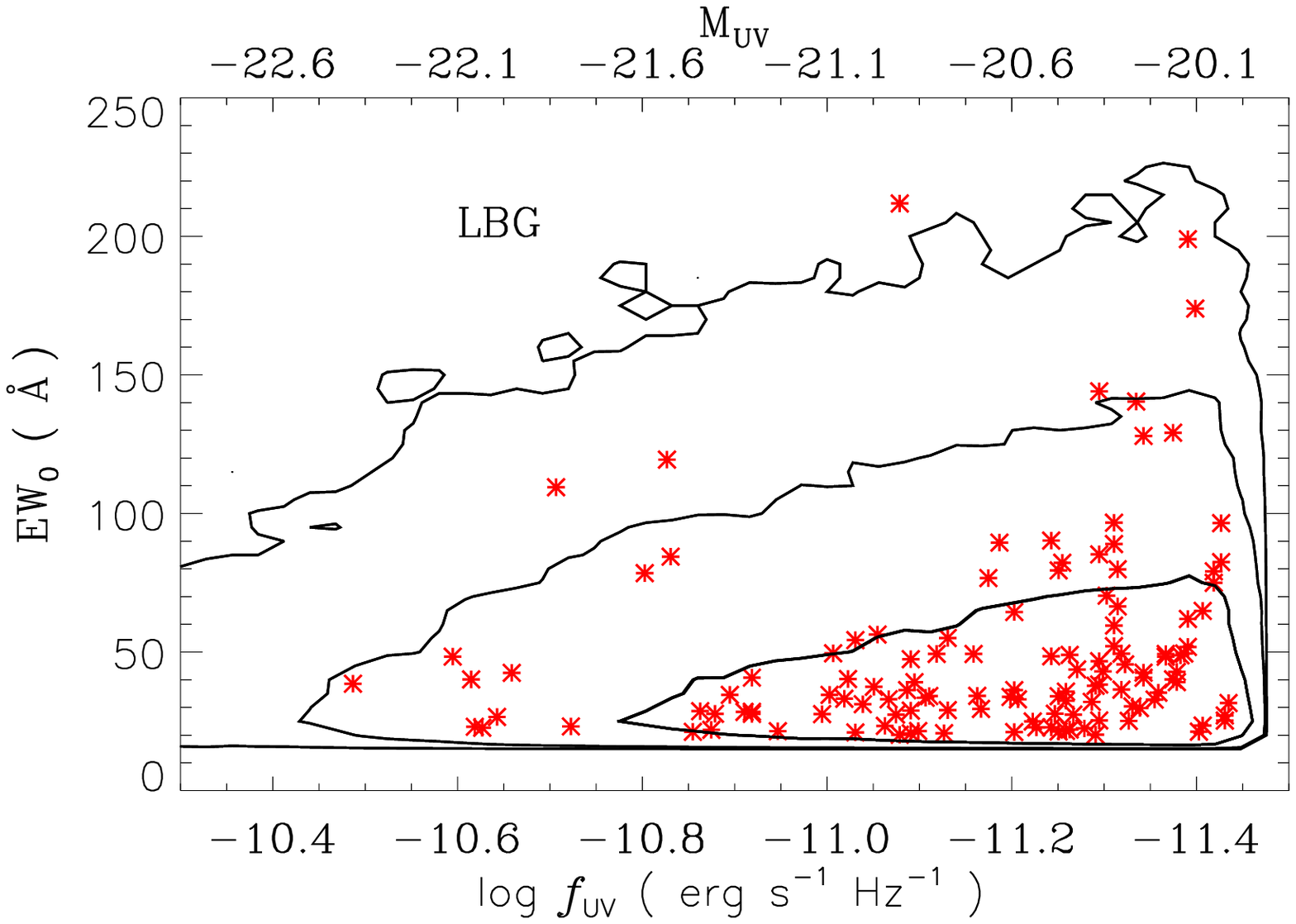,width=8.5cm}
 \caption{\emph{Left} Plot of EW versus $f_{UV}$ for the Ly$\alpha$ emitter sample. Black contours are 68\%, 95\% and 99.7\% confidence contours of the simulated test sample, with $u_{EW,LAE} = 0$, and the red points the observed points. The simulated sample is fully consistent with the observed sample. The lower limit to the distribution comes from the flux limitation in Ly$\alpha$. The upper left region of the plot is void of objects due to too small survey volumes. \emph{Right} The same plot for Lyman-break galaxies, with contours from the simulated sample with $u_{EW,LBG} =0$. }
 \label{fig:ando}
\end{figure*}

The random Ly$\alpha$ luminosities and EWs drawn for the sample of test galaxies can also be converted to the restframe UV absolute magnitude, $M_{UV}$, or vice versa. Thus, it is possible to reproduce the EW versus $M_{UV}$ plot (Ando et al. 2006) also for Ly$\alpha$ emitters. In Fig.~\ref{fig:ando} we show this plot for the Ly$\alpha$ emitter and Lyman-break galaxy samples and for $u_{EW} = 0$.
The region of high-EW, UV bright objects proposed by Ando et al. (2006) to lack objects was $M_{UV} < -21.5$ ($\log f_{UV} > -10.84$) and EW$_0 > 20$~{\AA}. As is seen in Fig.~\ref{fig:ando}, this apparent lack of objects is easily explained by small survey volumes. 
If there is indeed an independence between Ly$\alpha$ luminosity and EW, then in order to find an object with high-EW and bright UV emission, a galaxy has to be drawn from the low probability ends of two distributions. The probability to find such an object is thus very small.
In the example of $u_{EW} = 0$, and for galaxies with fluxes brighter than our simulation limits, the fraction of the total sample of galaxies in the quadrant with $M_{UV} < -21.5$ and $EW_0 > 20$~{\AA} is 24~\% for Ly$\alpha$ emitters and 11~\% for Lyman-break galaxies (with $M_{UV} < -20.0$ and EW$> 20$). The results of Ando et al. (2006) show two Ly$\alpha$ emitters in this quadrant from a total of eight and three Lyman-break galaxies out of 17 in the total sample. This corresponds to percentages $25\pm18$\% for Ly$\alpha$ emitters and $18\pm10$\% for Lyman-break galaxies, including very large error bars based on small number statistics, i.e. consistent with the simulated results. See also Table~\ref{tab:ando} for a summary on these results.
\begin{table}
 \centering
 \begin{minipage}{82mm}
  \caption{Percentages in high-EW, UV-bright corner in different samples. }
  \begin{tabular}{@{}lccccccccc@{}}
  \hline
   Type & Simulations ($u_{EW} =  0$)   &  Samples   &   Ando   &\\
 \hline
 LAE & 24\% & 31$\pm$4\% & 25$\pm$18\% &\\
 LBG & 11\% & 9$\pm$3\% & 18$\pm$10\% & \\
 \hline\label{tab:ando}
\end{tabular}
\begin{list}{}{}{}{}{}{}
\item[ ]  Percentages of objects in the quadrant with $M_{UV} < -21.5$ and EW$_0 > 20$~{\AA}. Second column gives the percentages in the simulated samples with $u_{EW} = 0$, third column in the observed samples used in this paper, and fourth column those of Ando et al. (2006). 
\end{list}
\end{minipage}
\end{table}
Based on the results for the sample from Shapley et al. (2003), there is a weak dependence between the EW and the UV flux, in the sense that there is a smaller fraction of high EW objects at brighter $M_{UV}$, but $u_{EW,LBG} = 0$ is only ruled out to $1.1\sigma$, or $73$\%. To confirm a real deficiency in objects in a quadrant would ideally require of the order 100 objects in that quadrant, in this case requiring a total sample of $> 500$ Ly$\alpha$ emitters or $> 1000$ Lyman-break galaxies with confirmed fluxes and EWs. It is important to note that the analysis performed here was made at redshift $z \sim 3$ as most of the data exists at this redshift. Samples at redshifts $z \sim 5 - 6$ are much smaller, and would produce a much weaker result. Several other factors make the higher redshift range more unsuitable for this test. Firstly, the measurement of the EW in very high redshift galaxies becomes very uncertain due to the increasingly faint UV continuum, in some cases resulting in lower limits on the measured EW. Secondly, all line emission surveys, that are not spectroscopically confirmed, contain some fraction of low redshift interlopers. This fraction is expected to increase with increasing redshift (c.f.~Kakazu et al.~2007). It is also uncertain if some neutral gas still remains at $z \sim 6$, which could affect the measurements of Ly$\alpha$ EW/flux in unpredictable ways. For the sample studied here, at $z\sim3$, these effects are not serious. Until larger samples have been collected at either redshift ranges, no conclusion can be drawn on whether there is a lack of high-EW, UV bright galaxies in the Universe.

\section{Discussion}\label{sec:discussion}
How can the $u_{EW}$ parameter be interpreted? A large positive $u_{EW}$ would indicate that there are fewer galaxies with large Ly$\alpha$ EWs at brighter UV fluxes. A large negative $u_{EW}$ indicates that it is more likely to find large-EW UV-bright galaxies. The next question is then what could cause this inflation or deflation of the EW distribution in the brighter UV flux slices? As Ando et al. (2006) argue, there could be several reasons for why there should be a dependence between these two measurable quantities. What is clear is that the time-scale on which the two quantities are sensitive to are different. The Ly$\alpha$ flux, and thus EW, is only large in the first $\sim 50$~Myrs, assuming a single stellar population. The UV flux on the other hand can stay large for several hundred Myrs. In each galaxy, the UV and Ly$\alpha$ flux is integrated over several star forming regions with different time-scales, with more massive galaxies having more star forming regions. If a larger fraction of these regions are in or out of their their Ly$\alpha$ emitting phase, this could cause the EW distribution to inflate or deflate for UV-bright galaxies.
Dust, metallicity, halo mass,
infall rate, star formation rate, gas dynamics, and detailed gas
morphology are also ingredients which could easily work to modify
$w_u$. 

In this paper we have searched for evidence of such an effect. We have
done this in two ways. First we used a simple null-hypothesis of
a single luminosity function and a constant $w_u$ for all luminosities.
For a well defined sample of 232 Ly$\alpha$ emitters, the null-hypothesis predicts that
56 of those should be in the (EW$ > 20$~{\AA}, $M_{UV} < -21.5$) section of
parameter space. We find 72, which is $2 \sigma$ more than predicted by
the simple model.
This result is at variance with earlier suggestions
that this section of the EW-$M_{UV}$ plane is under-populated, if anything
there might be marginal evidence that it is over-populated. Second,
in order to parametrize this analysis in a more global way, we
defined a description of the equivalent width distribution which
allows it to change as a function of Ly$\alpha$ luminosity via a parameter
$u_{EW,LAE}$. Following a Monte Carlo procedure we determined the best fit of
a large number of random realizations to our sample and found
$u_{EW,LAE} = -4^{+9}_{-7}$. This means that in a global sense the $u_{EW,LAE}$
parameter is consistent with zero for the entire sample, i.e.
consistent with a constant $w_u$. For the Lyman-break galaxies, the results are similar, with $u_{EW,LBG}$ consistent with zero within $1.1 \sigma$.

It is of interest to ask how strong the limits set by this result are.
We will use a simple comparison to set those limits into context. In a
recent paper (Nilsson et al.~2009) it was shown that $w_u$ is a function of
redshift and that it has been found to drop from $69$~{\AA} at $z = 3$ to
$48.5$~{\AA} at $z = 2.25$, i.e. a change of $20.5$~{\AA}. From Eq.~\ref{eq:ewlae} we see that  
the most extreme effect, reducing the $w_u$, occurs when $u_{EW,LAE}$ is as largely positive as possible. If we then include $u_{EW,LAE} = 5$, the upper envelope of the $1\sigma$ probability for the Ly$\alpha$ emitters, and compute what L$_{Ly\alpha}$ corresponds to a shift in $w_u = 20.5$~{\AA}, we find that this happens at a minimum Ly$\alpha$ luminosity of $10^{43.11}$~erg~s$^{-1}$, while 98.7\% of the sample have smaller luminosities than this. In conclusion, our analysis has constrained the dependence of $w_u$ on the Ly$\alpha$ luminosity to be smaller than, or of the order of, the change observed as a function of redshift.

The dependence-free case of $u_{EW} = 0$ found here is intriguing, as it means that the Ly$\alpha$ EW distribution is independent on the Ly$\alpha$ flux of Ly$\alpha$ emitters and the UV flux of Lyman-break galaxies, and is then independent to all parameters governing these fluxes (e.g. mass, age, metallicity, star formation rates etc.). In principle, the EW could still be correlated with the other respective parameter, i.e. the EW in Ly$\alpha$ emitters could be dependent on $M_{UV}$ and vice versa, although the dependency between Ly$\alpha$ flux and $M_{UV}$ in the galaxies would then have to be contrived. We consider such a scenario to be unlikely. 

The fact that our results are consistent with no dependence between Ly$\alpha$ EW and the UV flux means that effects of varying the IMF, dust and age are not very important over the flux range that we probe in present day surveys of high redshift galaxies. That there is no age dependence indicates that the light observed from these galaxies is integrated over many star forming regions with varying ages. The analysis presented here would greatly improve by adding more samples of both types of high redshift galaxies, but especially a larger sample of spectroscopically observed Lyman-break galaxies is necessary to finally draw any conclusions regarding the dependence between Ly$\alpha$ EW and UV flux in high redshift galaxies.

\section*{Acknowledgments}

The Dark Cosmology Centre is funded by the DNRF.

\end{document}